\documentclass[a4paper,12pt]{article}
\pdfoutput=1
\usepackage{bbm}
\usepackage{amsfonts,amsthm,amsmath,amssymb,latexsym,graphics,braket,color,ascmac,subfigure}
\usepackage[dvipdfm]{graphicx}
\usepackage{bm}
\usepackage{hyperref}
\usepackage{cite}

\renewcommand{\Re}{\operatorname{Re}}
\renewcommand{\Im}{\operatorname{Im}}

\begin{document}
\begin{titlepage}
\begin{flushright}
%{\small RIKEN-iTHEMS-Report-21, YITP-21-84}\\
{\small \today}
 \\
\end{flushright}

\begin{center}

\vspace{1cm}

\hspace{3mm}{\bf \Large Black Hole Singularity and Timelike Entanglement} \\[3pt] 

\vspace{1cm}

\renewcommand\thefootnote{\mbox{$\fnsymbol{footnote}$}}
Takanori Anegawa${}^{1,2}$\footnote{takanegawa@gmail.com}, and Kotaro {Tamaoka}${}^{3}$\footnote{tamaoka.kotaro@nihon-u.ac.jp}

\vspace{5mm}
${}^{1}${\small \sl Yonago College, National Institute of Technology
Yonago, Tottori 683-8502, Japan}\\
${}^{2}${\small \sl Department of Physics, Osaka University, Toyonaka, Osaka 560-0043, Japan}\\
${}^{3}${\small \sl Department of Physics, College of Humanities and Sciences, Nihon University, \\Sakura-josui, Tokyo 156-8550, Japan}\\

\end{center}

\vspace{5mm}

\noindent
\abstract
We study timelike and conventional entanglement entropy as potential probes of black hole singularities via the AdS/CFT correspondence. Using an analytically tractable example, we find characteristic behavior of holographic timelike entanglement entropy when the geometry involves a curvature singularity. We also observe interesting phenomena that, in some particular setups, holographic timelike and conventional entanglement entropy are determined from multiple complex saddle points, which fall outside the assumptions of the Lewkowycz-Maldacena type argument.

\end{titlepage}
\setcounter{footnote}{0}
\renewcommand\thefootnote{\mbox{\arabic{footnote}}}
%%%%%%%%%%%%%%%TITLE%%%%%%%%%%%%%%%%\newpage
\tableofcontents
\flushbottom
%\newpage
\section{Introduction and Summary}\label{sec:intro}
Understanding the quantum nature of gravity has progressed significantly through studying black holes. The AdS/CFT correspondence~\cite{Maldacena:1997re}, an ideal framework for exploring the holographic principle, suggests that the issues in semi-classical black holes can be resolved once full non-perturbative quantum (and stringy) effects are taken into account. In this paper, we are interested in the curvature singularities inside black holes and how signals of such singularities can be read off from boundary quantities via the AdS/CFT correspondence.

%This paper addresses an important issue in classical black holes: the curvature singularity inside the black hole. %Ultimately, we aim to understand what (wrong) procedure gives rise to the problems in semi-classical black holes and then how to resolve them using the non-perturbative framework as the AdS/CFT correspondence. 

In the context of AdS/CFT correspondence, the emergence of bulk spacetime is closely related to quantum entanglement at the boundary~\cite{Maldacena:2001kr, Ryu:2006bv, Swingle:2009bg,VanRaamsdonk:2010pw}. The entanglement entropy, the unique measure of bi-partite entanglement for pure states, can capture the spatial structure of spacetime through the spacelike extremal surfaces in the bulk~\cite{Ryu:2006bv,Hubeny:2007xt}. On the other hand, timelike surfaces are irrelevant to the entanglement entropy and other well-studied information theoretic quantities such as (holographic) complexity~\cite{Susskind:2014rva}.  Since the radial direction inside a black hole, where the singularity exists, shifts from a spatial to a temporal direction, understanding timelike surfaces and their boundary counterparts is expected to be crucial for comprehending spacetime singularities via holography.

Recently, it has become possible to understand some of these timelike surfaces through a quantity called timelike entanglement entropy~\cite{Doi:2022iyj,Doi:2023zaf}, which can be interpreted as a special situation of pseudo entropy~\cite{Nakata:2020luh}. Let two states $\ket{\psi}, \ket{\varphi}$ belong to a bipartite Hilbert space $\mathcal{H}_A\otimes\mathcal{H}_{B}$. Then, the pseudo entropy is defined as von Neumann entropy of the reduced transition matrix, 
\begin{align}
S_A^{(P)}=\textrm{Tr}_A\left[-\tau_A\log\tau_A\right],
\end{align}
where the (reduced) transition matrix is given by
\begin{align}
\tau_A&=\textrm{Tr}_{B}\left[\tau\right],\\
\tau&=\dfrac{\ket{\psi}\hspace{-1mm}\bra{\varphi}}{\braket{\varphi|\psi}}.
\end{align}
The pseudo entropy can be interpreted as a generalization of entanglement entropy to the post-selection process. Refer to information theoretical interpretation of the pseudo entropy~\cite{Nakata:2020luh,Akal:2021dqt} and further developments~\cite{Mollabashi:2021xsd,Miyaji:2021lcq,Mollabashi:2020yie,Goto:2021kln,Nishioka:2021cxe,Camilo:2021dtt,Murciano:2021dga,Ishiyama:2022odv,Guo:2022sfl,Mukherjee:2022jac,Bhattacharya:2022wlp,Guo:2022jzs,Alshal:2023kcd,He:2023wko,Kawamoto:2023nki,Parzygnat:2023avh,Chen:2023gnh,Omidi:2023env,He:2023syy,Kanda:2023jyi,Guo:2023tjv,Shinmyo:2023eci,He:2023eap,Guo:2023aio,He:2024jog,Guo:2024edr,Guo:2024wmj}. We can show that holographic pseudo entropy can be calculated as the area of the minimal surface in a Euclidean time-dependent asymptotically AdS background\footnote{For later convenience, we note that this statement is verified via a Lewkowycz-Maldacena type argument~\cite{Lewkowycz:2013nqa}, meaning that we assumed a single saddle point on the gravity side as like the Ryu-Takayanagi formula.}, similar to Ryu-Takayanagi formula~\cite{Ryu:2006bv}. Then, the timelike entanglement entropy $S^{(T)}_A$ can be defined as the analytic continuation of spacelike subsystem $A$ to a timelike one. See also \cite{Reddy:2022zgu,Narayan:2022afv,Li:2022tsv,Foligno:2023dih,Jiang:2023ffu,Chu:2023zah,Jiang:2023loq,He:2023ubi,Das:2023yyl,Narayan:2023zen,Narayan:2023ebn,Guo:2024lrr,Grieninger:2023knz,Basu:2024bal,Afrasiar:2024lsi,Carignano:2024jxb} for subsequent work on timelike entanglement. %timelike entanglement entropy is used to discuss the relationship with the emergence of time in the dS/CFT correspondence~\cite{Doi:2022iyj} and suggests a novel type of holographic prescription within the AdS/CFT framework~\cite{Doi:2023zaf}. 

In this paper, we investigate the holographic timelike entanglement associated with black holes with curvature singularities to determine the extent to which timelike entanglement entropy can capture the physics of singularities. In previous studies for timelike entanglement entropy, the discussion was limited to black holes without curvature singularities. 
In Section \ref{sec:oneside} and \ref{sec:twoside}, we will employ rather traditional approaches for calculating minimal surfaces to address this problem and discuss the connection between our approaches and previous studies (see Section \ref{subsec:htee}). The similar attempts using the correlation functions have been discussed in \cite{Kraus:2002iv,Fidkowski:2003nf,Festuccia:2005pi,Grinberg:2020fdj,Rodriguez-Gomez:2021pfh,deBoer:2022zps,Horowitz:2023ury,Ceplak:2024bja,Singhi:2024sdr}, for example. 

In Section \ref{sec:oneside}, we discuss timelike subsystems defined on a single boundary of AdS- Schwarzschild-type black holes. We found that the presence of the curvature singularity causes the spatial surfaces to approach almost null, unlike in the BTZ black hole, where there is no linear growth in time $t$. Notably, in the model considered, there is no imaginary part up to the multivaluedness of the logarithm. 

In Section \ref{sec:twoside}, we move to subsystems defined on both two boundaries. We argue that the multiple complex saddle points will contribute simultaneously to timelike entanglement, otherwise it negatively diverges. This must be required because the setup can be also identified with the usual entanglement entropy for the Hartman-Maldacena setup~\cite{Hartman:2013qma}, making it an intriguing example that contradicts the assumptions of the Lewkowycz-Maldacena type argument~\cite{Lewkowycz:2013nqa}.

In Section \ref{sec:discussion}, we conclude with discussions. In particular, we propose an intriguing possibility that we need to introduce the end of the world brane around the black hole singularity based on UV/UV relation inside of the horizon. 

In Appendix \ref{app:time}, we leave a technical part of the analytic continuation of boundary subsystems for the black hole with the curvature singularity. 

\section{Single-sided setup and complex geodesics}\label{sec:oneside}
\subsection{Preliminary}\label{subsec:setup}
\begin{figure}[t]
    \centering
    \subfigure[]{\includegraphics[width=7cm]{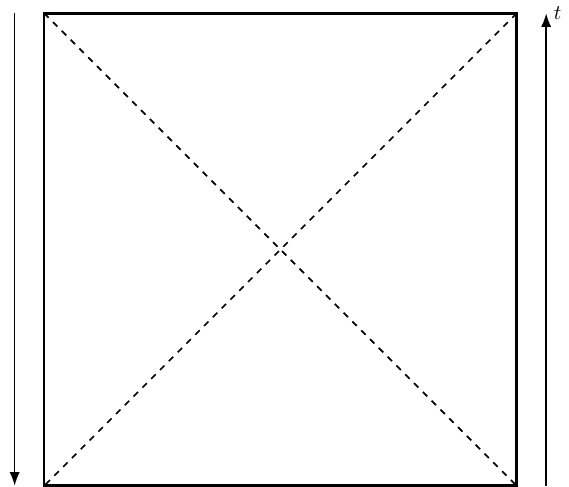}}\hspace{1cm}
    \subfigure[]{\includegraphics[width=7cm]{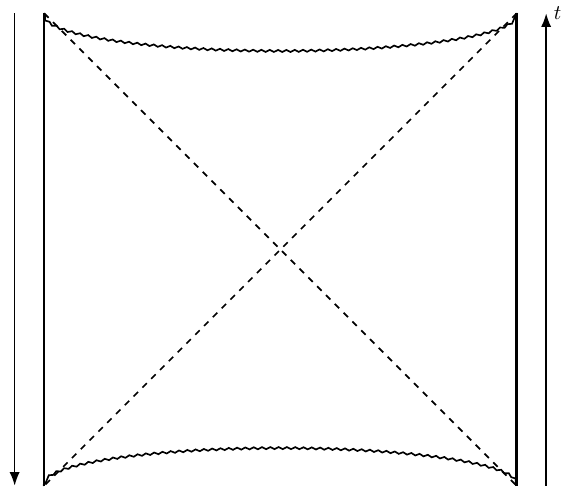}}
    \caption{Penrose diagrams for (a) BTZ black hole and (b) AdS-Schwarzschild black hole. The crucial difference between them is the presence of a curvature singularity at the origin $r=0$ (wavy lines). Another important characteristic of the AdS-Schwarzschild black hole is that the $r=0$ surface concaves the Penrose diagram~\cite{Fidkowski:2003nf}. This difference in the global structure also affects the structure of geodesics discussed later.}
    \label{fig:black hole}
\end{figure}

Throughout this paper, we
consider three-dimensional asymptotically AdS spacetime of the following metric
\begin{align}
    ds^2&=-f(r)dt^2+\dfrac{dr^2}{f(r)}+r^2dx^2. \label{eq:bh}
\end{align}
To be more specific, we will discuss the two cases:
\begin{align}
f(r)=\begin{cases}
    r^2-r_+^2& (\text{BTZ black hole}),\\
    r^2-\dfrac{r_+^4}{r^2}& (\text{AdS-Schwarzschild-like black hole}),
\end{cases}
\end{align}
where $r_+$ is the radius of the event horizon for each geometry\footnote{Later, we will often set $r_+=1$ for simplicity. Then, Hawking temperature $\beta=4\pi/|f^\prime(r_+)|$ becomes $2\pi$ for BTZ black hole and $\pi$ for AdS-Schwarzschild-like black hole. }. 
The first case is the BTZ black hole which is a vacuum solution of pure AdS${}_3$ Einstein gravity and can be understood as an orbifold of pure AdS${}_3$~\cite{Banados:1992wn,Banados:1992gq}. Namely, the BTZ black hole has a singularity at $r=0$, but only a conical one. The second case is a black hole with curvature singularity. We would like to treat these geometries as examples of black holes with and without curvature singularity. See Figure \ref{fig:black hole} for Penrose diagrams of these geometries.  

The second geometry is somewhat unusual in three-dimension, so we elaborate on it in what follows. We can interpret this geometry as the dimensional reduction of planer AdS-Schwarzschild black hole in five-dimension. Therefore, this geometry should be understood as a solution of Einstein equation with certain matter fields in light of three-dimension. Since some ambiguities remain in the holographic interpretation of timelike entanglement entropy in higher dimensions~\cite{Doi:2023zaf}, we will stick to three-dimensional spacetime to avoid unnecessary subtleties.

To discuss the timelike entanglement entropy, we first compute the pseudo-entropy for the Wick-rotated geometry, 
\begin{align}
    ds^2&=f(r)d\tau^2+\dfrac{dr^2}{f(r)}+r^2dx^2. \label{eq:bh_euclidean}
\end{align}
We would like to first calculate the holographic pseudo entropy~\cite{Nakata:2020luh} for the interval $A=\{-T/2\leq\tau\leq T/2\}$, and then obtain the timelike entanglement entropy by the analytic continuation $T\rightarrow iT$. 
To this end, we minimize the area functional defined as
\begin{align}
A=\int d\lambda \sqrt{f(r)\dot{\tau}^2+f(r)^{-1}\dot{r}^2},
\end{align}
where $\dot{\tau}=\frac{d\tau}{d\lambda}$ and $\dot{r}=\frac{dr}{d\lambda}$. We choose $\lambda$ such that
\begin{align}
1=f(r)\dot{\tau}^2+\dfrac{\dot{r}^2}{f(r)}. \label{eq:proper_length}
\end{align}
Since the metric \eqref{eq:bh_euclidean} has translational symmetry along Euclidean time $\tau$, we have the conserved energy defined as 
\begin{align}
f(r)\dot{\tau}=\tilde{E}. \label{eq:energy_euclidean}
\end{align}
For later convenience, we also note an integral expression of $\tau(\lambda)$ that is equivalent to \eqref{eq:energy_euclidean},
\begin{align}
    \tau(\lambda)=\tilde{E}\int\dfrac{d\lambda}{f(r(\lambda))}. \label{eq:time_euclidean}
\end{align}
Combining \eqref{eq:proper_length} and \eqref{eq:energy_euclidean}, we obtain
\begin{align}
    A_{\text{min.}}(\tilde{E})=2\int^{r_b}_{r_c} \dfrac{dr}{\sqrt{f(r)-\tilde{E}^2}},
\end{align}
where
$r_b$ is the boundary cutoff surface taken to be large and $r_c$ is a turning point defined as
\begin{align}
\dot{r}^2=f(r_c)-\tilde{E}^2=0 .\label{eq:mech}
\end{align}
The location of the turning point (after the analytic continuation) will give how deeper bulk regime we can probe via the minimal area surface. As we will see later, large $\tilde{E}$ corresponds to geodesics probing only close to the asymptotic boundary, while small $\tilde{E}$ corresponds to ones going through the horizon. 

\subsection{BTZ black hole}\label{subsec:btz}
We first discuss the case of BTZ black hole as a warm-up exercise. The timelike entanglement entropy, in this case, has been already studied in \cite{Doi:2022iyj,Doi:2023zaf}. The purpose of this subsection is to present another (but rather traditional) approach that will be useful for the AdS-Schwarzschild-like case later. In particular, we argue the connection between the present approach and the holographic proposal in \cite{Doi:2023zaf}. 

Solving the equation of motion, we obtain the radial trajectory as
\begin{align}
r(\lambda)=\sqrt{r_+^2+\tilde{E}^2}\cosh(\lambda-\lambda_0), \label{eq:rad_btz}
\end{align}
where $\lambda_0$ corresponds to the proper length at the critical value $\dot{r}(\lambda_0)=0$. By inverting the above expression, we obtain the minimal area $A_{\text{min.}}$ as
\begin{align}
A_{\text{min.}}(\tilde{E})=2[\lambda(r=r_b)-\lambda_0]=2\cosh^{-1}\left[\dfrac{r_b}{\sqrt{r_+^2+\tilde{E}^2}}\right]\simeq 2\log\dfrac{2r_b}{\sqrt{r_+^2+\tilde{E}^2}} \label{eq:area_btz}
\end{align}
In the last equality, we assume the cutoff surface $r=r_b$ to be large and keep the leading order of large $r_b$ expansion. 
We would like to translate the minimal surface as a function of the energy into a function of Euclidean time. To this end, let us express the time separation $T=\Delta \tau$ at the boundary as a function of energy, 
\begin{align}
\Delta \tau (\tilde{E})=\lim_{\lambda\rightarrow\infty}[\tau(\lambda)-\tau(-\lambda)]= \dfrac{2}{r_+}\tan^{-1}\left[\dfrac{r_+}{\tilde{E}}\right],\label{eq:etime_btz}
\end{align}
where
\begin{align}
\tau (\lambda)=\dfrac{1}{r_+}\tan^{-1}\left[\dfrac{r_+}{\tilde{E}}\tanh(\lambda-\lambda_0)\right].\label{eq:etime_btz2}
\end{align}
Combining \eqref{eq:area_btz} and \eqref{eq:etime_btz}, we obtain
\begin{align}
A_{\text{min.}}\simeq 2\log\left[\dfrac{2r_b}{r_+}\sin\dfrac{r_+}{2}T\right].
\end{align}
After the analytic continuation $T\rightarrow iT$, we obtain the timelike entanglement as
\begin{align}
S^{(T)}_A=\dfrac{A_{\text{min.}}}{4G_N}=\dfrac{c}{3}\log\left[\dfrac{2r_b}{r_+}\sinh\dfrac{r_+}{2}T\right]+\dfrac{c\pi}{6}i. \label{eq:tee_btz1}
\end{align}
This exactly matches the CFT and holographic result derived in \cite{Doi:2022iyj,Doi:2023zaf}. 
\begin{figure}[htbp]
    \centering
    \subfigure[]{\includegraphics[width=6.8cm]{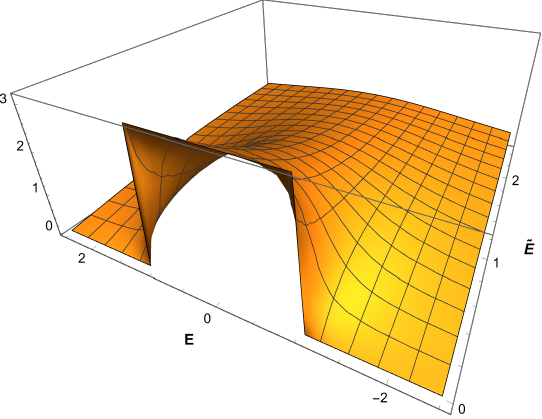}}\hspace{1cm}
    \subfigure[]{\includegraphics[width=6.8cm]{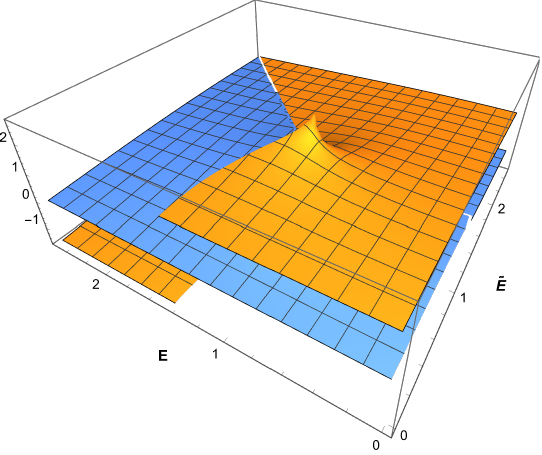}}
    \caption{3d plots of $\Re[\Delta \tau(\tilde{E}+iE)]$ for (a) BTZ black hole (first-sheet) and (b) AdS-Schwarzschild-like black hole (first and second sheet). In BTZ case, the real part are just divided into $|E|<1$ and others, where $\Re [\Delta \tau(iE)] = \pi (=\beta/2)$ if $|E|<1$ and $\Re[\Delta \tau(iE)] = 0$ if $|E|>1$. In Schwarzschild-like case, a simple analytic continuation (starting from a particular value $\tilde{E}$, and rotation to imaginary axis) can end up in the range of orange sheet (first sheet) or move beyond the branch to blue sheet (second sheet). This gives $\Re[\Delta \tau(iE)] = 0$ if $E$ is above a certain value $E_{\rm cr}$, and $\Re[\Delta \tau(iE)] = \pi/2 (=\beta/2)$ if $E$ is below a certain value $E_{\rm cr}$.}
    \label{fig:re_time}
\end{figure}

\begin{figure}[htbp]
    \centering
    \subfigure[]{\includegraphics[width=7cm]{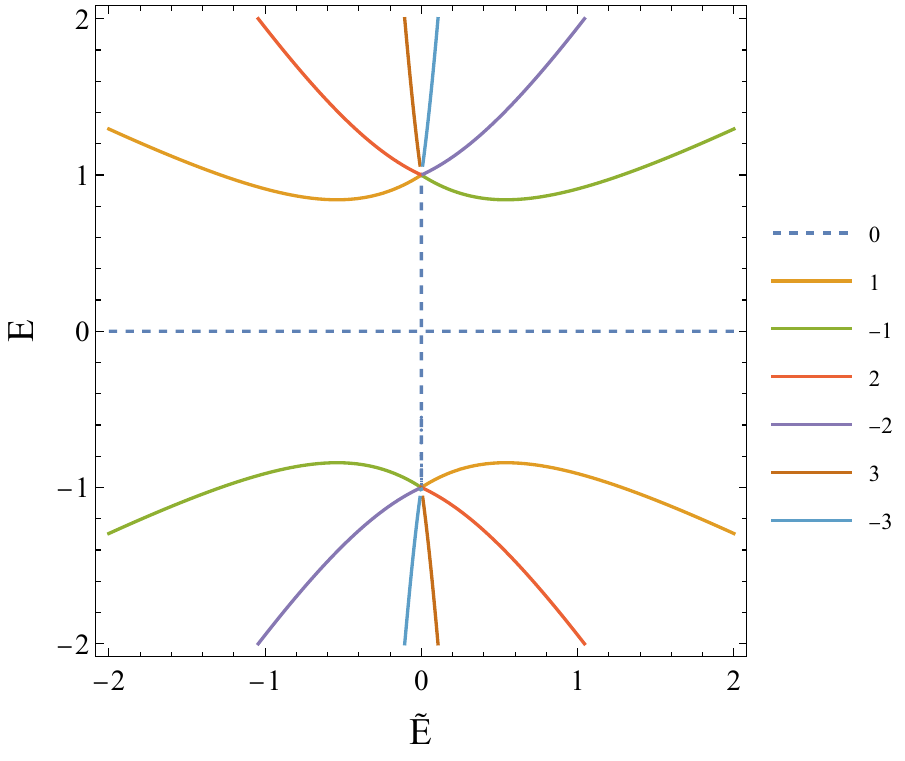}}\hspace{1cm}
    \subfigure[]{\includegraphics[width=7cm]{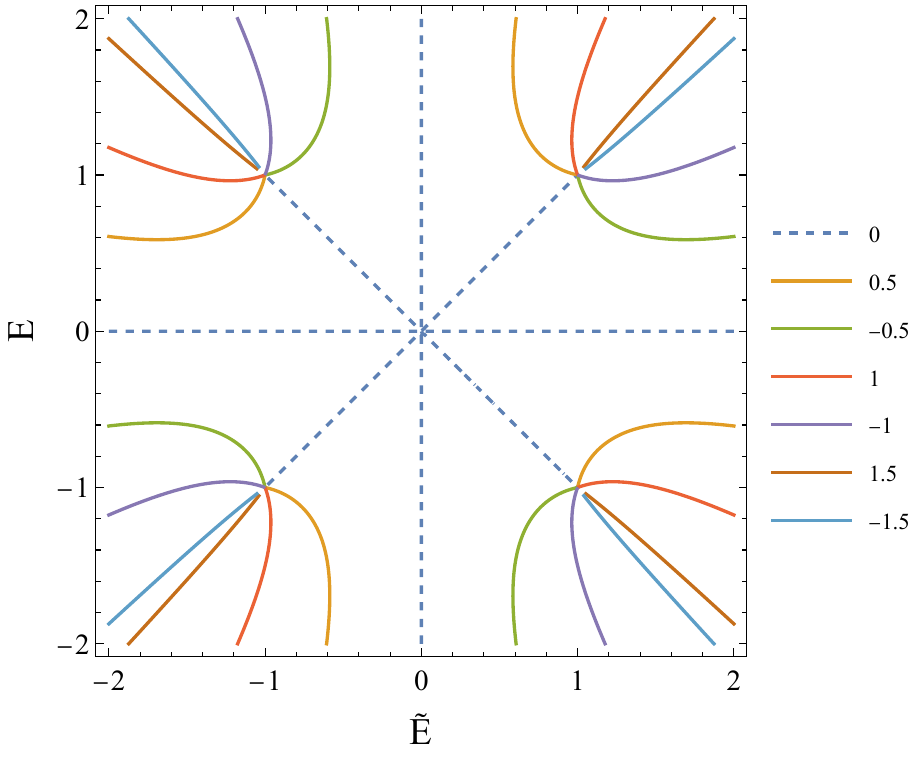}}
    \caption{Contour plots of $\Im[A_{\text{min.}}(\tilde{E}+iE)]$ for (a) BTZ black hole and (b) AdS-Schwarzschild-like black hole. Each dashed line corresponds to $\Im[A_{\text{min.}}(\tilde{E}+iE)]=0$, hence expected to be no contribution from the timelike geodesics. }
    \label{fig:im_area}
\end{figure}
Note that the above analytic continuation is not unique in terms of energy as the area $A_{\text{min.}}(iE)$ and time interval $\Delta\tau(iE)$ has logarithmic branch cuts along $E>1$ and $E<-1$. If we assumed $\tilde{E}<1$ and rotated to the Lorentzian energy $E<1$, we would have $\Re \Delta\tau(iE)=\beta/2$, thus the resulting geodesics is anchored on different sides of the maximally extended black hole geometry. Such analytic continuation corresponds to the Hartman-Maldacena type surface~\cite{Hartman:2013qma} and gives the real-valued area. Since we are now interested in the single-sided setup, we would like to keep $\Re \Delta\tau(iE)=0$. To this end, we need to assume $\tilde{E}>1$ and rotate to the Lorentzian energy $E>1$. This gives complex geodesics and results in the complex-valued area as \eqref{eq:tee_btz1}. See also left panel of Figure \ref{fig:re_time} and \ref{fig:im_area}.

\subsection{Connection to the holographic prescription}\label{subsec:htee}
\begin{figure}[t]
    \centering
    \includegraphics[width=7cm]{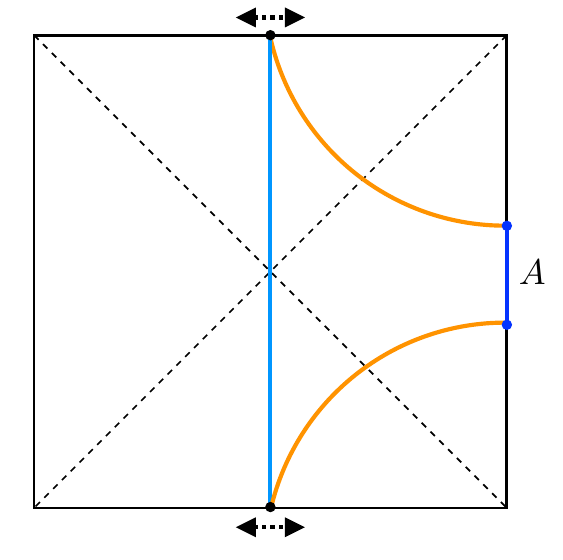}
    \caption{Schematic picture of holographic prescription proposed in \cite{Doi:2023zaf}. The candidates of geodesics $\Gamma_A$ consist of a union of spacelike (orange curves) and timelike (blue curve in the center) geodesics such that $\partial\Gamma_A=\partial A$ as like the standard extremal surface. Then, we consider variation with respect to the joining points (black dots around the center). We require they are all stationary.}
    \label{fig:htee}
\end{figure}

In \cite{Doi:2023zaf}, the authors have proposed a holographic prescription to obtain the timelike entanglement entropy discussed above. Interestingly, this prescription involves a novel type of minimal surface that consists of some union of spacelike and timelike surfaces. See Figure \ref{fig:htee}.

Here we would like to ask about the connection between our (rather traditional) calculation and their proposal. 
From \eqref{eq:rad_btz}, \eqref{eq:etime_btz} and \eqref{eq:etime_btz2}, we obtain the radial and temporal trajectory as
\begin{align}
r(\lambda)^2&=\dfrac{r_+^2}{\sin^2\frac{r_+T}{2}}\cosh^2(\lambda-\lambda_0),\\
\tau(\lambda)&=\dfrac{1}{r_+}\tan^{-1}\left[\tan\left(\frac{r_+T}{2}\right)\tanh(\lambda-\lambda_0)\right].
\end{align}
After analytic continuation, this may look different from the holographic picture discussed in \cite{Doi:2022iyj,Doi:2023zaf} because the proposed minimal surfaces involve spacelike surfaces that end on the conical singularity at $r=0$. 

To consult with their holographic prescription, we should notice that the area is now complex-valued. To get the geometrical interpretation, we need to notice that the proper length (namely, the affine parameter) should be real valued. For this purpose, we subtract the imaginary part $i\pi$ from the analytically continued area $A_{\text{min.}}$. This subtraction corresponds to replacing the proper length $\lambda$ with $\lambda-i\frac{\pi}{2}$. The analytic continuation $T\rightarrow iT$ together with this replacement results in 
\begin{align}
r(\lambda)^2&=\dfrac{r_+^2}{\sinh^2\frac{r_+T}{2}}\sinh^2(\lambda-\lambda_0), \label{eq:realpart_btz}\\
t(\lambda)&=\dfrac{1}{r_+}\tanh^{-1}\left[\tanh\left(\frac{r_+T}{2}\right)\coth(\lambda-\lambda_0)\right]. \label{eq:realpart_btz2}
\end{align}
Notice that the radial trajectory reaches the singularity $r=0$ at the turning point. The temporal trajectory \eqref{eq:realpart_btz2} also ensures that inside the horizon it is a timelike surface and consistent with Figure \ref{fig:htee}. We also stress that the above expression \eqref{eq:realpart_btz} correctly reproduces the real part of $A_{\text{min.}}$. It is worth noting that if we put $r_+=i$, we can repeat the similar argument for the timelike entanglement in the case of global AdS${}_3$. 
\subsection{AdS-Schwarzschild-like black hole}\label{subsec:adssch}

As mentioned earlier, we would like to focus on the three-dimensional black hole solution and dual two-dimensional conformal field theory. This motivates us to consider the above-mentioned
dimensionally reduced AdS-Schwarzschild-like black hole metric,
\begin{align}
    f(r)=r^2-\dfrac{1}{r^2}. \label{eq:f_ads_sch}
\end{align}
We can easily solve the equation \eqref{eq:mech} for \eqref{eq:f_ads_sch}. That turns out to be
\begin{align}
   r(\lambda)^2=\frac{1}{2}\sqrt{\tilde{E}^4 + 4} \cosh(2(\lambda - \lambda_0)) + \frac{\tilde{E}^2}{2}. 
\end{align}
In particular, the critical point $r_c$ is determined by
\begin{align}
2r_c^2=\tilde{E}^2+\sqrt{\tilde{E}^4+4}. \label{eq:critical_adssch}
\end{align}
It is worth noting that after analytic continuation, $\tilde{E}\rightarrow iE$, the critical point reaches the singularity as we increase the energy $E\gg1$. Therefore, when approaching the curvature singularity, our minimal surface approaches almost null,
\begin{align}
A_{\text{min.}}(\tilde{E})=\cosh^{-1}\left[\dfrac{2r_b^2-\tilde{E}^2}{\sqrt{4+\tilde{E}^4}}\right]\simeq \log\dfrac{4r_b^2}{\sqrt{4+\tilde{E}^4}}. \label{eq:area_sch}
\end{align}
%\begin{align}
%    \Delta \tau_E = \tan^{-1}\left( \frac{\sqrt{\tilde{E}^4 + 4} - \tilde{E}^2 + 2}{2\tilde{E}} \right)-\tanh^{-1} \left( \frac{\sqrt{\tilde{E}^4 + 4} - \tilde{E}^2 - 2}{2\tilde{E}} \right) \label{eq:etime_adssch2}
%\end{align}
In the last expression, we assumed $r_b\gg\tilde{E}$ so that the $r=r_b$ surface works as a cutoff surface. 
On the other hand, the boundary Euclidean time interval is computed as
\begin{align}
    \Delta \tau(\tilde{E}) = \dfrac{1}{2}\log\left[\dfrac{\tilde{E}^2+2\tilde{E}+2}{\sqrt{\tilde{E}^4+4}}\right]-\dfrac{i}{2}\log\left[\dfrac{\tilde{E}^2+2i\tilde{E}-2}{\sqrt{\tilde{E}^4+4}}\right].\label{eq:etime_adssch1}
\end{align}
Similar to the BTZ black hole, the small energy limit $\tilde{E}\ll1$  corresponds to $\Delta \tau(\tilde{E})=\pi/2=\beta/2$ ({\it i.e.} the spacelike geodesics anchored on two different asymptotic boundaries), while the large energy limit $\tilde{E}\gg1$ corresponds to $\Delta \tau(\tilde{E})=0$. 

Since the boundary time interval $\Delta\tau(\tilde{E})$ involves logarithmic branch cuts, we have no unique choice of the analytic continuation with respect to the complex energy $\tilde{E}+iE$. Here we want to consider the interval $A$ situated only on the single-sided boundary, so we need to choose the one that eventually leads to the trivial Euclidean time separation $\Re\Delta\tau(iE)=0$. To this end, we expand $\Delta\tau(\tilde{E})$ around $\tilde{E}\gg1$ regime, 
\begin{align}
\Delta\tau(\tilde{E})=\dfrac{2}{\tilde{E}}+\mathcal{O}(\tilde{E}^{-5}),
\end{align}
and then Wick rotate to the Lorentzian energy $\tilde{E}\rightarrow iE$ (see also right panel of Figure \ref{fig:re_time}). After this analytic continuation, we obtain %$\Delta \tau \rightarrow -iT$ (or $\tilde{E}\rightarrow iE$ with large $\tilde{E}$), 
\begin{align}
A_{\text{min.}}\simeq \log\dfrac{4r_b^2}{\sqrt{\tilde{E}^4}}\simeq\dfrac{1}{2}\log \left[16r_b^4\left(\dfrac{\Delta\tau}{2}\right)^4\right]\rightarrow \dfrac{1}{2}\left(\log r_b^4T^4-2\pi i\right).
\end{align}
At this moment, we postpone determining the branch cut of the logarithm and keep $-2\pi i$ explicitly. It is worth noting that we have no ``lightcone singularity'', except for the zero size interval limit ($T\rightarrow0$). The absence of such singularity is in contrast with the two-sided cases discussed in the next section. This difference indeed comes from a different choice for the analytic continuation of $\Delta\tau(\tilde{E})$. See appendix \ref{app:time} for more details. 

Interestingly, we would have no imaginary part in the final expression if we fixed a branch for the logarithm. This is also natural from the point of view of the full expression \eqref{eq:critical_adssch}. After fixing the branch of logarithm from $-\pi$ to $\pi$, we have
\begin{align}
    S^{(T)}_A=\dfrac{A_{\text{min.}}}{4G_N}\simeq\dfrac{c}{3}\log r_bT,
\end{align}
where we used the relation between the central charge and the Newton constant $c=3/2G_N$~\cite{Brown:1986nw}. It means that our geodesics are still real (see also right panel of Figure \ref{fig:im_area}), and the critical point \eqref{eq:critical_adssch} reaches to the curvature singularity as $E\rightarrow\infty$.

In contrast with the BTZ black hole, {\it i.e.,} a black hole with no curvature singularity, we need infinite energy to get close to the origin $r=0$\footnote{Note, however, that the minimal surface for the BTZ black hole never approaches to the black hole singularity if we do not apply the interpretation discussed in Section \ref{subsec:htee}. These are qualitatively different in any interpretation.}. If we take a smaller energy (a larger time interval), the minimal surface gradually moves away from the singularity. This behavior is qualitatively different from the BTZ black hole discussed in the previous section, suggesting that timelike entanglement can tell us the structure of black hole singularity. 

While the timelike entanglement entropy for our setup nicely captures the information of singularity, we have no imaginary part for its principal value. It implies that our geometrical picture looks slightly different from that of the holographic prescription for the BTZ black hole in Figure \ref{fig:htee}. We will discuss potential resolutions in Section \ref{sec:discussion}. 
\section{Comments on two-sided setup}\label{sec:twoside}
\begin{figure}[t]
    \centering
    \includegraphics[width=6cm]{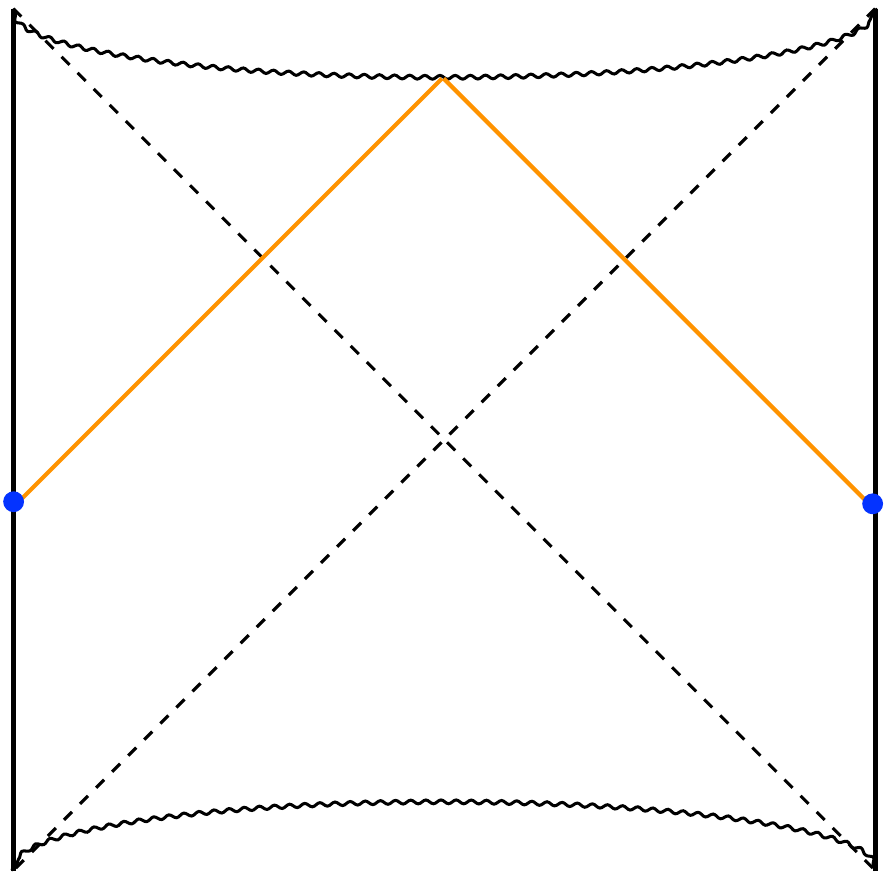}
    \caption{Holographic entanglement entropy (or timelike entanglement entropy) for a two-sided setup. }
    \label{fig:ads-sch2}
\end{figure}
So far we have studied timelike entanglement entropy for a single interval located on a single-side boundary of black holes. Here we continue our study to a two-side setup, $A=\{-t_L<t, t_R<t\}$, where $t_L$ and $t_R$ represent the Killing time for the left and right boundaries. In this case, we want to keep Euclidean time interval $\Re\Delta\tau=\beta/2$ and this condition corresponds to expand Euclidean energy of $\tau(\tilde{E})$ around $\tilde{E}=0$. 

Since we now consider infinite size systems, timelike entanglement entropy should be equivalent to the entanglement entropy for a semi-infinite interval $A^\prime=\{x_R>0, x_L>0\}$. This type of entropy can be calculated as the length of geodesics anchored on two different sides of the boundaries~\cite{Hartman:2013qma},
\begin{align}
    S_A^{(T)}=S_{A^{\prime}}=\dfrac{c}{3}\log\left[\dfrac{r_b \beta}{\pi}\cosh\dfrac{2\pi t}{\beta}\right].
\end{align}
At the late time, the entropy growth linearly in time, $S_{A^{\prime}}\propto t$, representing the growth of wormhole length. 

Next, we consider the case of the AdS-Schwarzschild-like black hole. The similar calculation of real geodesics anchored on two different boundaries~\cite{Fidkowski:2003nf} gives rise to
\begin{align}
A_{\text{min.}}=\log\dfrac{4r_b^2}{\sqrt{4+E^4}}= \log r_b^2(t-t_c)^2+\cdots,
\end{align}
where  $t_c=-\pi/2$. Therefore, we have negative divergence around $t=t_c$ due to the geodesics bouncing off the curvature singularity (see Figure \ref{fig:ads-sch2}). 
If we regard it as the pseudo entropy, this behavior might be allowed as pseudo entropy has no upper/lower bound in principle\footnote{This negative divergence is not like the amplification similar to the weak value discussed in \cite{Ishiyama:2022odv} since we are picking up an eigenvalue of ``area operator'' associated with a single geometry. In this case, we never encounter such amplification phenomena.}. However, since we can interpret this configuration as the usual entanglement entropy, this behavior must be problematic. 

 Essentially the same problem and its resolution have been pointed out in the context of boundary two-point function and bulk-bulk propagators in ~\cite{Fidkowski:2003nf,Festuccia:2005pi}. We can understand what was wrong with the previous calculation from the Euclidean energy and time as follows. If we expand energy around $\tilde{E}=0$, which is necessary to obtain the two-sided geodesics, we have
\begin{align}
    \Delta\tau(\tilde{E})&=\dfrac{\pi}{2}-\dfrac{\tilde{E}^3}{6}+\mathcal{O}(\tilde{E}^{7}),\\
    A_{\min}(\tilde{E})&=\log 2r_b^2-\dfrac{\tilde{E}^4}{8}+\mathcal{O}(\tilde{E}^{8}).
\end{align}
It means that we have a branch point singularity at $\Delta\tau=\pi/2$ of the form $(\Delta\tau-\pi/2)^{\frac{4}{3}}$.
Therefore, we have three possible configurations in terms of Euclidean energy. One of them corresponds to the real geodesics related to the above-mentioned one after the analytic continuation. The remaining two geodesics are complex-valued and conjugate to each other. It turns out that the minimal geodesic is not the real one, but the pair of complex geodesics. This situation is out of the standard assumption for calculating the holographic entanglement entropy, where we assume the dominant saddle point~\cite{Lewkowycz:2013nqa}. Therefore, we cannot give a naive geometrical interpretation in this setup.  

\section{Discussion}\label{sec:discussion}
The timelike entanglement entropy of the AdS-Schwarzschild-like black hole calculated in Section \ref{subsec:adssch} does not include the imaginary part. This is also evident from the expression of pseudo entropy discussed around \eqref{eq:area_sch}. Naively, this appears to contradict the proposal for timelike entanglement entropy in \cite{Doi:2023zaf}. We would like to emphasize again that the solution we have dealt with here is not a solution in pure AdS gravity. Therefore, the reason for the discrepancy might be that we did not take the contribution from matter fields seriously. We also have to suspect that it originates from similar ambiguities in higher-dimensional examples~\cite{Doi:2023zaf}. We leave this issue as important future work. In what follows, we propose another interesting possibility. 
\subsection*{End of the world brane and final state projection}
\begin{figure}[t]
    \centering
    \includegraphics[width=7cm]{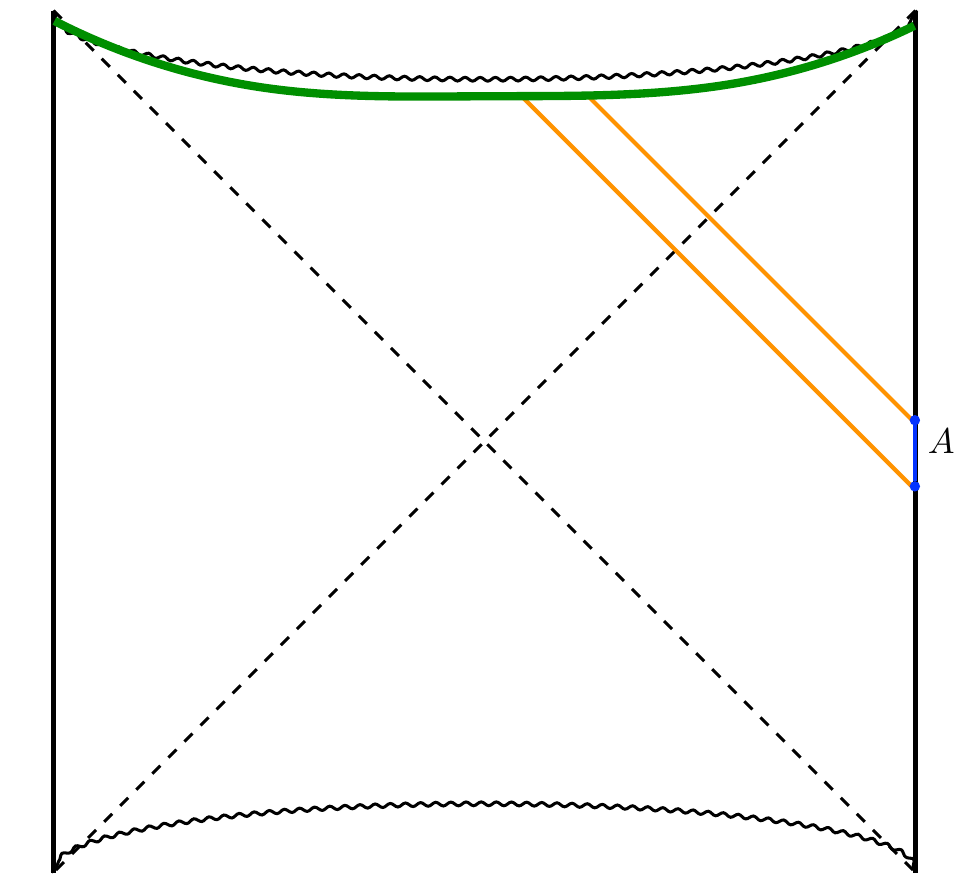}
    \caption{If we introduce the end of the world brane floating on $r=r_c$ (green curve), we can split the spacelike geodesics into two pieces (orange lines), hence we need not include timelike geodesics.}
    \label{fig:etw}
\end{figure}

As is clear from the definition of $r=r_c$ surface, inside a black hole, the narrower the width of the boundary system, equivalently the higher the energy resolution, the closer to singularity the region can be probed. In other words, the UV/IR relation in the standard AdS/CFT correspondence now becomes a UV/UV relation inside the horizon as pointed out in \cite{Festuccia:2005pi}. Therefore, it is natural to expect that the corresponding conditions are imposed even inside the horizon since the boundary side involves UV cutoff and boundary conditions on the entangling surface. In other words, we can expect the end of the world brane to exist on the $r=r_c$ surface. If this expectation is correct, it justifies the absence of imaginary parts. That is, two spacelike surfaces can be separated along the end of the world brane (see Figure \ref{fig:etw}). This might be understood as a realization of the final state projection~\cite{Horowitz:2003he}.

\subsection*{Complex saddle points in holographic entanglement entropy}

In the two-sided setup, we discussed the contribution of multiple complex saddle points; from the calculation of holographic entanglement entropy, assuming that only saddle points contribute, no other combination of these two saddles was possible. While it is challenging to take the path integral of gravity seriously, it is known that the real saddle point (even if it is minimal) cannot contribute if one thinks of it as the WKB approximation of the two-point function of a scalar field in the bulk~\cite{Festuccia:2005pi}. Although holographic entanglement entropy is expected to take a real saddle point in the usual pure AdS gravity, one should keep in mind the possibility that multiple complex saddle points may contribute. 

If we consider the spacetime considered in this paper as a dimensional reduction of the five-dimensional solution, a similar divergence cannot occur in the calculation of the real-valued extremal surfaces in the higher dimension. It would be interesting to see if the present calculations when properly incorporating the matter fields, can be interpreted within the original black hole geometry and avoid pathological behavior like the island formula~\cite{Penington:2019npb,Almheiri:2019psf,Almheiri:2019hni}.

\section*{Acknowledgement}
We thank Shunichiro Kinoshita for fruitful discussions. We also thank Tomonori Ugajin and Tadashi Takayanagi for useful comments on the draft. K.T.~is supported by JSPS KAKENHI Grant No.~21K13920 and MEXT KAKENHI Grant No.~24H00972. 

\appendix
\section{Analytic continuation for the single-sided intervals}\label{app:time}

\begin{figure}[t]
    \centering
    \includegraphics[width=7cm]{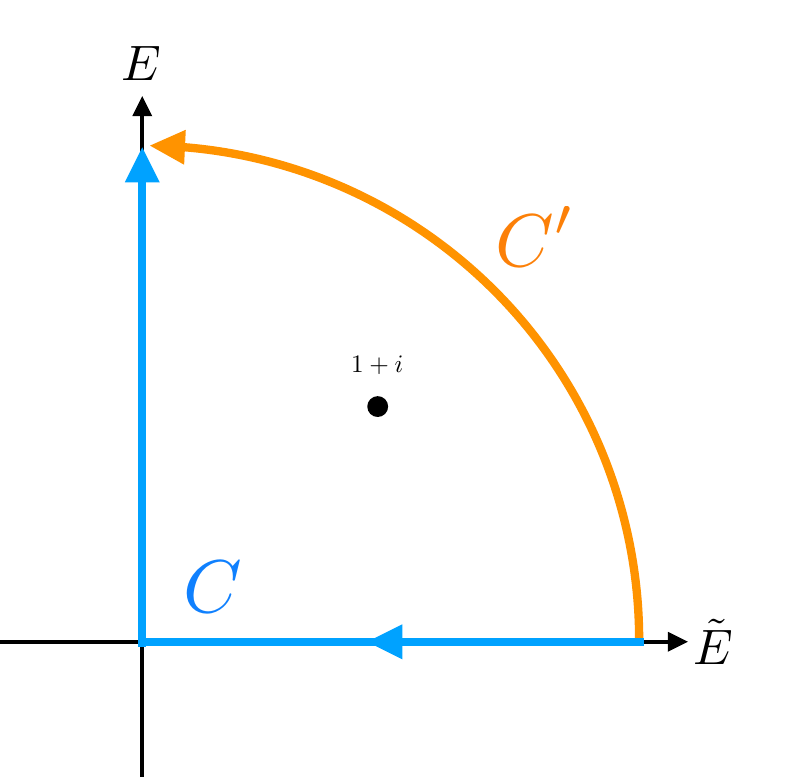}
    \caption{Contours for two different setups. We use $C$ (blue curve) for the two-sided setup, while $C^\prime$ (orange curve) for the one-sided setup. These differences can be represented as the residue (from a black dot inside of $C^\prime-C$) of \eqref{eq:deriv_tau}. }
    \label{fig:contours}
\end{figure}

In the main text, we used the analytic continuation of time interval function $\Delta\tau$ that is different from the one for two-sided geodesics often discussed in the literature. We first recast the expression of $\Delta \tau (\tilde{E})$ for convenience, 
\begin{align}
    \Delta \tau (\tilde{E}) = \dfrac{1}{2}\log\left[\dfrac{\tilde{E}^2+2\tilde{E}+2}{\sqrt{\tilde{E}^4+4}}\right]-\dfrac{i}{2}\log\left[\dfrac{\tilde{E}^2+2i\tilde{E}-2}{\sqrt{\tilde{E}^4+4}}\right].
\end{align}
We can simply use $\Delta \tau (iE)$ if we are interested in the two-sided setup as $\Re[\Delta \tau (iE)]=\beta/2$. To demonstrate the different choices of the analytic continuation explicitly, we first rewrite the above expression in a contour integral as,
\begin{align}
    \Delta \tau_C(E)\equiv\Delta \tau (iE)=\int_C dz\left(-\dfrac{2z^2}{z^4+4}\right),
\end{align}
where the contour $C$ starts at infinity on the real axis, runs along the real axis to the origin, and then to the large value $E$ on the imaginary axis. See Figure \ref{fig:contours}. Here the integrand is simply the derivative of $\tau (\tilde{E})$,
\begin{align}
\dfrac{d\Delta \tau (\tilde{E})}{d\tilde{E}}=-\dfrac{2\tilde{E}^2}{\tilde{E}^4+4}. \label{eq:deriv_tau}
\end{align}
In particular, these functions involve the singularity at $\tilde{E}=\sqrt{2}\mathrm{e}^{i\frac{1+2k}{4}\pi},\,(k=0,1,2,3)$. 
Let us take a different path $C^\prime$ (see Figure \ref{fig:contours}) that corresponds to Wick rotation to the Lorentzian energy $E\rightarrow\mathrm{e}^{i\frac{\pi}{2}}E$. Then, the difference of these choices is simply expressed as the residue of the pole at $1+i$,
\begin{align}
\Delta \tau_{C^\prime}(E)-\Delta \tau_C(E)=2\pi i\,\textrm{Res}\left(\frac{d\Delta \tau (\tilde{E})}{d\tilde{E}},1+i\right)=-\dfrac{i\pi}{2}-\dfrac{\pi}{2}.
\end{align}
In Section \ref{sec:oneside}, we used $\Delta \tau_{C^\prime}(E)$ rather than $\Delta \tau_{C}(E)$. 
The shift for the Euclidean time ensures that our geodesics are anchored on the single asymptotic boundary, while the shift for the Lorentzian time resolves the light-cone singularity discussed in Section \ref{sec:twoside}.  
\bibliographystyle{JHEP}
\bibliography{reference.bib}

\providecommand{\href}[2]{#2}\begingroup\raggedright\begin{thebibliography}{10}

\bibitem{Maldacena:1997re}
J.~M. Maldacena, \emph{{The Large N limit of superconformal field theories and
  supergravity}}, \href{https://doi.org/10.4310/ATMP.1998.v2.n2.a1}{\emph{Adv.
  Theor. Math. Phys.} {\bfseries 2} (1998) 231}
  [\href{https://arxiv.org/abs/hep-th/9711200}{{\ttfamily hep-th/9711200}}].

\bibitem{Maldacena:2001kr}
J.~M. Maldacena, \emph{{Eternal black holes in anti-de Sitter}},
  \href{https://doi.org/10.1088/1126-6708/2003/04/021}{\emph{JHEP} {\bfseries
  04} (2003) 021} [\href{https://arxiv.org/abs/hep-th/0106112}{{\ttfamily
  hep-th/0106112}}].

\bibitem{Ryu:2006bv}
S.~Ryu and T.~Takayanagi, \emph{{Holographic derivation of entanglement entropy
  from AdS/CFT}},
  \href{https://doi.org/10.1103/PhysRevLett.96.181602}{\emph{Phys. Rev. Lett.}
  {\bfseries 96} (2006) 181602}
  [\href{https://arxiv.org/abs/hep-th/0603001}{{\ttfamily hep-th/0603001}}].

\bibitem{Swingle:2009bg}
B.~Swingle, \emph{{Entanglement Renormalization and Holography}},
  \href{https://doi.org/10.1103/PhysRevD.86.065007}{\emph{Phys. Rev. D}
  {\bfseries 86} (2012) 065007}
  [\href{https://arxiv.org/abs/0905.1317}{{\ttfamily 0905.1317}}].

\bibitem{VanRaamsdonk:2010pw}
M.~Van~Raamsdonk, \emph{{Building up spacetime with quantum entanglement}},
  \href{https://doi.org/10.1142/S0218271810018529}{\emph{Gen. Rel. Grav.}
  {\bfseries 42} (2010) 2323}
  [\href{https://arxiv.org/abs/1005.3035}{{\ttfamily 1005.3035}}].

\bibitem{Hubeny:2007xt}
V.~E. Hubeny, M.~Rangamani and T.~Takayanagi, \emph{{A Covariant holographic
  entanglement entropy proposal}},
  \href{https://doi.org/10.1088/1126-6708/2007/07/062}{\emph{JHEP} {\bfseries
  07} (2007) 062} [\href{https://arxiv.org/abs/0705.0016}{{\ttfamily
  0705.0016}}].

\bibitem{Susskind:2014rva}
L.~Susskind, \emph{{Computational Complexity and Black Hole Horizons}},
  \href{https://doi.org/10.1002/prop.201500092}{\emph{Fortsch. Phys.}
  {\bfseries 64} (2016) 24} [\href{https://arxiv.org/abs/1403.5695}{{\ttfamily
  1403.5695}}].

\bibitem{Doi:2022iyj}
K.~Doi, J.~Harper, A.~Mollabashi, T.~Takayanagi and Y.~Taki,
  \emph{{Pseudoentropy in dS/CFT and Timelike Entanglement Entropy}},
  \href{https://doi.org/10.1103/PhysRevLett.130.031601}{\emph{Phys. Rev. Lett.}
  {\bfseries 130} (2023) 031601}
  [\href{https://arxiv.org/abs/2210.09457}{{\ttfamily 2210.09457}}].

\bibitem{Doi:2023zaf}
K.~Doi, J.~Harper, A.~Mollabashi, T.~Takayanagi and Y.~Taki, \emph{{Timelike
  entanglement entropy}},
  \href{https://doi.org/10.1007/JHEP05(2023)052}{\emph{JHEP} {\bfseries 05}
  (2023) 052} [\href{https://arxiv.org/abs/2302.11695}{{\ttfamily
  2302.11695}}].

\bibitem{Nakata:2020luh}
Y.~Nakata, T.~Takayanagi, Y.~Taki, K.~Tamaoka and Z.~Wei, \emph{{New
  holographic generalization of entanglement entropy}},
  \href{https://doi.org/10.1103/PhysRevD.103.026005}{\emph{Phys. Rev. D}
  {\bfseries 103} (2021) 026005}
  [\href{https://arxiv.org/abs/2005.13801}{{\ttfamily 2005.13801}}].

\bibitem{Akal:2021dqt}
I.~Akal, T.~Kawamoto, S.-M. Ruan, T.~Takayanagi and Z.~Wei, \emph{{Page curve
  under final state projection}},
  \href{https://doi.org/10.1103/PhysRevD.105.126026}{\emph{Phys. Rev. D}
  {\bfseries 105} (2022) 126026}
  [\href{https://arxiv.org/abs/2112.08433}{{\ttfamily 2112.08433}}].

\bibitem{Mollabashi:2021xsd}
A.~Mollabashi, N.~Shiba, T.~Takayanagi, K.~Tamaoka and Z.~Wei, \emph{{Aspects
  of pseudoentropy in field theories}},
  \href{https://doi.org/10.1103/PhysRevResearch.3.033254}{\emph{Phys. Rev.
  Res.} {\bfseries 3} (2021) 033254}
  [\href{https://arxiv.org/abs/2106.03118}{{\ttfamily 2106.03118}}].

\bibitem{Miyaji:2021lcq}
M.~Miyaji, \emph{{Island for gravitationally prepared state and pseudo
  entanglement wedge}},
  \href{https://doi.org/10.1007/JHEP12(2021)013}{\emph{JHEP} {\bfseries 12}
  (2021) 013} [\href{https://arxiv.org/abs/2109.03830}{{\ttfamily
  2109.03830}}].

\bibitem{Mollabashi:2020yie}
A.~Mollabashi, N.~Shiba, T.~Takayanagi, K.~Tamaoka and Z.~Wei, \emph{{Pseudo
  Entropy in Free Quantum Field Theories}},
  \href{https://doi.org/10.1103/PhysRevLett.126.081601}{\emph{Phys. Rev. Lett.}
  {\bfseries 126} (2021) 081601}
  [\href{https://arxiv.org/abs/2011.09648}{{\ttfamily 2011.09648}}].

\bibitem{Goto:2021kln}
K.~Goto, M.~Nozaki and K.~Tamaoka, \emph{{Subregion spectrum form factor via
  pseudoentropy}},
  \href{https://doi.org/10.1103/PhysRevD.104.L121902}{\emph{Phys. Rev. D}
  {\bfseries 104} (2021) L121902}
  [\href{https://arxiv.org/abs/2109.00372}{{\ttfamily 2109.00372}}].

\bibitem{Nishioka:2021cxe}
T.~Nishioka, T.~Takayanagi and Y.~Taki, \emph{{Topological pseudo entropy}},
  \href{https://doi.org/10.1007/JHEP09(2021)015}{\emph{JHEP} {\bfseries 09}
  (2021) 015} [\href{https://arxiv.org/abs/2107.01797}{{\ttfamily
  2107.01797}}].

\bibitem{Camilo:2021dtt}
G.~Camilo and A.~Prudenziati, \emph{{Twist operators and pseudo entropies in
  two-dimensional momentum space}},
  \href{https://arxiv.org/abs/2101.02093}{{\ttfamily 2101.02093}}.

\bibitem{Murciano:2021dga}
S.~Murciano, P.~Calabrese and R.~M. Konik, \emph{{Generalized entanglement
  entropies in two-dimensional conformal field theory}},
  \href{https://doi.org/10.1007/JHEP05(2022)152}{\emph{JHEP} {\bfseries 05}
  (2022) 152} [\href{https://arxiv.org/abs/2112.09000}{{\ttfamily
  2112.09000}}].

\bibitem{Ishiyama:2022odv}
Y.~Ishiyama, R.~Kojima, S.~Matsui and K.~Tamaoka, \emph{{Notes on pseudo
  entropy amplification}},
  \href{https://doi.org/10.1093/ptep/ptac112}{\emph{PTEP} {\bfseries 2022}
  (2022) 093B10} [\href{https://arxiv.org/abs/2206.14551}{{\ttfamily
  2206.14551}}].

\bibitem{Guo:2022sfl}
W.-z. Guo, S.~He and Y.-X. Zhang, \emph{{On the real-time evolution of
  pseudo-entropy in 2d CFTs}},
  \href{https://doi.org/10.1007/JHEP09(2022)094}{\emph{JHEP} {\bfseries 09}
  (2022) 094} [\href{https://arxiv.org/abs/2206.11818}{{\ttfamily
  2206.11818}}].

\bibitem{Mukherjee:2022jac}
J.~Mukherjee, \emph{{Pseudo Entropy in U(1) gauge theory}},
  \href{https://doi.org/10.1007/JHEP10(2022)016}{\emph{JHEP} {\bfseries 10}
  (2022) 016} [\href{https://arxiv.org/abs/2205.08179}{{\ttfamily
  2205.08179}}].

\bibitem{Bhattacharya:2022wlp}
A.~Bhattacharya, A.~Bhattacharyya and S.~Maulik, \emph{{Pseudocomplexity of
  purification for free scalar field theories}},
  \href{https://doi.org/10.1103/PhysRevD.106.086010}{\emph{Phys. Rev. D}
  {\bfseries 106} (2022) 086010}
  [\href{https://arxiv.org/abs/2209.00049}{{\ttfamily 2209.00049}}].

\bibitem{Guo:2022jzs}
W.-z. Guo, S.~He and Y.-X. Zhang, \emph{{Constructible reality condition of
  pseudo entropy via pseudo-Hermiticity}},
  \href{https://doi.org/10.1007/JHEP05(2023)021}{\emph{JHEP} {\bfseries 05}
  (2023) 021} [\href{https://arxiv.org/abs/2209.07308}{{\ttfamily
  2209.07308}}].

\bibitem{Alshal:2023kcd}
H.~Alshal, \emph{{Einstein\textquoteright{}s equations and the pseudo-entropy
  of pseudo-Riemannian information manifolds}},
  \href{https://doi.org/10.1007/s10714-023-03130-7}{\emph{Gen. Rel. Grav.}
  {\bfseries 55} (2023) 86} [\href{https://arxiv.org/abs/2301.13017}{{\ttfamily
  2301.13017}}].

\bibitem{He:2023wko}
S.~He, J.~Yang, Y.-X. Zhang and Z.-X. Zhao, \emph{{Pseudo entropy of primary
  operators in $ T\overline{T}/J\overline{T} $-deformed CFTs}},
  \href{https://doi.org/10.1007/JHEP09(2023)025}{\emph{JHEP} {\bfseries 09}
  (2023) 025} [\href{https://arxiv.org/abs/2305.10984}{{\ttfamily
  2305.10984}}].

\bibitem{Kawamoto:2023nki}
T.~Kawamoto, S.-M. Ruan, Y.-k. Suzuki and T.~Takayanagi, \emph{{A half de
  Sitter holography}},
  \href{https://doi.org/10.1007/JHEP10(2023)137}{\emph{JHEP} {\bfseries 10}
  (2023) 137} [\href{https://arxiv.org/abs/2306.07575}{{\ttfamily
  2306.07575}}].

\bibitem{Parzygnat:2023avh}
A.~J. Parzygnat, T.~Takayanagi, Y.~Taki and Z.~Wei, \emph{{SVD entanglement
  entropy}}, \href{https://doi.org/10.1007/JHEP12(2023)123}{\emph{JHEP}
  {\bfseries 12} (2023) 123}
  [\href{https://arxiv.org/abs/2307.06531}{{\ttfamily 2307.06531}}].

\bibitem{Chen:2023gnh}
Z.~Chen, \emph{{Complex-valued Holographic Pseudo Entropy via Real-time AdS/CFT
  Correspondence}},  \href{https://arxiv.org/abs/2302.14303}{{\ttfamily
  2302.14303}}.

\bibitem{Omidi:2023env}
F.~Omidi, \emph{{Pseudo R\'enyi Entanglement Entropies For an Excited State and
  Its Time Evolution in a 2D CFT}},
  \href{https://arxiv.org/abs/2309.04112}{{\ttfamily 2309.04112}}.

\bibitem{He:2023syy}
S.~He, Y.-X. Zhang, L.~Zhao and Z.-X. Zhao, \emph{{Entanglement and Pseudo
  Entanglement Dynamics versus Fusion in CFT}},
  \href{https://arxiv.org/abs/2312.02679}{{\ttfamily 2312.02679}}.

\bibitem{Kanda:2023jyi}
H.~Kanda, T.~Kawamoto, Y.-k. Suzuki, T.~Takayanagi, K.~Tasuki and Z.~Wei,
  \emph{{Entanglement phase transition in holographic pseudo entropy}},
  \href{https://doi.org/10.1007/JHEP03(2024)060}{\emph{JHEP} {\bfseries 03}
  (2024) 060} [\href{https://arxiv.org/abs/2311.13201}{{\ttfamily
  2311.13201}}].

\bibitem{Guo:2023tjv}
W.-z. Guo, Y.-z. Jiang and Y.~Jiang, \emph{{Pseudo entropy and
  pseudo-Hermiticity in quantum field theories}},
  \href{https://doi.org/10.1007/JHEP05(2024)071}{\emph{JHEP} {\bfseries 05}
  (2024) 071} [\href{https://arxiv.org/abs/2311.01045}{{\ttfamily
  2311.01045}}].

\bibitem{Shinmyo:2023eci}
K.~Shinmyo, T.~Takayanagi and K.~Tasuki, \emph{{Pseudo entropy under joining
  local quenches}}, \href{https://doi.org/10.1007/JHEP02(2024)111}{\emph{JHEP}
  {\bfseries 02} (2024) 111}
  [\href{https://arxiv.org/abs/2310.12542}{{\ttfamily 2310.12542}}].

\bibitem{He:2023eap}
S.~He, J.~Yang, Y.-X. Zhang and Z.-X. Zhao, \emph{{Pseudoentropy for descendant
  operators in two-dimensional conformal field theories}},
  \href{https://doi.org/10.1103/PhysRevD.109.025014}{\emph{Phys. Rev. D}
  {\bfseries 109} (2024) 025014}
  [\href{https://arxiv.org/abs/2301.04891}{{\ttfamily 2301.04891}}].

\bibitem{Guo:2023aio}
W.-z. Guo and J.~Zhang, \emph{{Sum rule for the pseudo-R\'enyi entropy}},
  \href{https://doi.org/10.1103/PhysRevD.109.106008}{\emph{Phys. Rev. D}
  {\bfseries 109} (2024) 106008}
  [\href{https://arxiv.org/abs/2308.05261}{{\ttfamily 2308.05261}}].

\bibitem{He:2024jog}
S.~He, P.~H.~C. Lau and L.~Zhao, \emph{{Detecting quantum chaos via
  pseudo-entropy and negativity}},
  \href{https://arxiv.org/abs/2403.05875}{{\ttfamily 2403.05875}}.

\bibitem{Guo:2024edr}
W.-z. Guo, Y.-z. Jiang and J.~Xu, \emph{{Pseudoentropy sum rule by analytical
  continuation of the superposition parameter}},
  \href{https://arxiv.org/abs/2405.09745}{{\ttfamily 2405.09745}}.

\bibitem{Guo:2024wmj}
W.-z. Guo and T.~Liu, \emph{{Non-Hermitian spacetime and generalized
  thermofield double formalism}},
  \href{https://arxiv.org/abs/2406.06961}{{\ttfamily 2406.06961}}.

\bibitem{Lewkowycz:2013nqa}
A.~Lewkowycz and J.~Maldacena, \emph{{Generalized gravitational entropy}},
  \href{https://doi.org/10.1007/JHEP08(2013)090}{\emph{JHEP} {\bfseries 08}
  (2013) 090} [\href{https://arxiv.org/abs/1304.4926}{{\ttfamily 1304.4926}}].

\bibitem{Reddy:2022zgu}
K.~S. Reddy, \emph{{A timelike entangled island at the initial singularity in a
  JT FLRW ($\Lambda>0$) universe}},
  \href{https://arxiv.org/abs/2211.14893}{{\ttfamily 2211.14893}}.

\bibitem{Narayan:2022afv}
K.~Narayan, \emph{{de Sitter space, extremal surfaces, and time entanglement}},
  \href{https://doi.org/10.1103/PhysRevD.107.126004}{\emph{Phys. Rev. D}
  {\bfseries 107} (2023) 126004}
  [\href{https://arxiv.org/abs/2210.12963}{{\ttfamily 2210.12963}}].

\bibitem{Li:2022tsv}
Z.~Li, Z.-Q. Xiao and R.-Q. Yang, \emph{{On holographic time-like entanglement
  entropy}}, \href{https://doi.org/10.1007/JHEP04(2023)004}{\emph{JHEP}
  {\bfseries 04} (2023) 004}
  [\href{https://arxiv.org/abs/2211.14883}{{\ttfamily 2211.14883}}].

\bibitem{Foligno:2023dih}
A.~Foligno, T.~Zhou and B.~Bertini, \emph{{Temporal Entanglement in Chaotic
  Quantum Circuits}},
  \href{https://doi.org/10.1103/PhysRevX.13.041008}{\emph{Phys. Rev. X}
  {\bfseries 13} (2023) 041008}
  [\href{https://arxiv.org/abs/2302.08502}{{\ttfamily 2302.08502}}].

\bibitem{Jiang:2023ffu}
X.~Jiang, P.~Wang, H.~Wu and H.~Yang, \emph{{Timelike entanglement entropy and
  TT\textasciimacron{} deformation}},
  \href{https://doi.org/10.1103/PhysRevD.108.046004}{\emph{Phys. Rev. D}
  {\bfseries 108} (2023) 046004}
  [\href{https://arxiv.org/abs/2302.13872}{{\ttfamily 2302.13872}}].

\bibitem{Chu:2023zah}
C.-S. Chu and H.~Parihar, \emph{{Time-like entanglement entropy in AdS/BCFT}},
  \href{https://doi.org/10.1007/JHEP06(2023)173}{\emph{JHEP} {\bfseries 06}
  (2023) 173} [\href{https://arxiv.org/abs/2304.10907}{{\ttfamily
  2304.10907}}].

\bibitem{Jiang:2023loq}
X.~Jiang, P.~Wang, H.~Wu and H.~Yang, \emph{{Timelike entanglement entropy in
  dS$_{3}$/CFT$_{2}$}},
  \href{https://doi.org/10.1007/JHEP08(2023)216}{\emph{JHEP} {\bfseries 08}
  (2023) 216} [\href{https://arxiv.org/abs/2304.10376}{{\ttfamily
  2304.10376}}].

\bibitem{He:2023ubi}
P.-Z. He and H.-Q. Zhang, \emph{{Timelike Entanglement Entropy from Rindler
  Method}},  \href{https://arxiv.org/abs/2307.09803}{{\ttfamily 2307.09803}}.

\bibitem{Das:2023yyl}
A.~Das, S.~Sachdeva and D.~Sarkar, \emph{{Bulk reconstruction using timelike
  entanglement in (A)dS}},
  \href{https://doi.org/10.1103/PhysRevD.109.066007}{\emph{Phys. Rev. D}
  {\bfseries 109} (2024) 066007}
  [\href{https://arxiv.org/abs/2312.16056}{{\ttfamily 2312.16056}}].

\bibitem{Narayan:2023zen}
K.~Narayan, \emph{{Further remarks on de Sitter space, extremal surfaces, and
  time entanglement}},
  \href{https://doi.org/10.1103/PhysRevD.109.086009}{\emph{Phys. Rev. D}
  {\bfseries 109} (2024) 086009}
  [\href{https://arxiv.org/abs/2310.00320}{{\ttfamily 2310.00320}}].

\bibitem{Narayan:2023ebn}
K.~Narayan and H.~K. Saini, \emph{{Notes on time entanglement and
  pseudo-entropy}},
  \href{https://doi.org/10.1140/epjc/s10052-024-12855-x}{\emph{Eur. Phys. J. C}
  {\bfseries 84} (2024) 499}
  [\href{https://arxiv.org/abs/2303.01307}{{\ttfamily 2303.01307}}].

\bibitem{Guo:2024lrr}
W.-z. Guo, S.~He and Y.-X. Zhang, \emph{{Relation between timelike and
  spacelike entanglement entropy}},
  \href{https://arxiv.org/abs/2402.00268}{{\ttfamily 2402.00268}}.

\bibitem{Grieninger:2023knz}
S.~Grieninger, K.~Ikeda and D.~E. Kharzeev, \emph{{Temporal entanglement
  entropy as a probe of renormalization group flow}},
  \href{https://doi.org/10.1007/JHEP05(2024)030}{\emph{JHEP} {\bfseries 05}
  (2024) 030} [\href{https://arxiv.org/abs/2312.08534}{{\ttfamily
  2312.08534}}].

\bibitem{Basu:2024bal}
D.~Basu and V.~Raj, \emph{{Reflected entropy and timelike entanglement in
  $\textrm{T}\bar{\textrm{T}}$ deformed CFT$_2$s}},
  \href{https://arxiv.org/abs/2402.07253}{{\ttfamily 2402.07253}}.

\bibitem{Afrasiar:2024lsi}
M.~Afrasiar, J.~K. Basak and D.~Giataganas, \emph{{Timelike Entanglement
  Entropy and Phase Transitions in non-Conformal Theories}},
  \href{https://arxiv.org/abs/2404.01393}{{\ttfamily 2404.01393}}.

\bibitem{Carignano:2024jxb}
S.~Carignano and L.~Tagliacozzo, \emph{{Loschmidt echo, emerging dual unitarity
  and scaling of generalized temporal entropies after quenches to the critical
  point}},  \href{https://arxiv.org/abs/2405.14706}{{\ttfamily 2405.14706}}.

\bibitem{Kraus:2002iv}
P.~Kraus, H.~Ooguri and S.~Shenker, \emph{{Inside the horizon with AdS / CFT}},
  \href{https://doi.org/10.1103/PhysRevD.67.124022}{\emph{Phys. Rev. D}
  {\bfseries 67} (2003) 124022}
  [\href{https://arxiv.org/abs/hep-th/0212277}{{\ttfamily hep-th/0212277}}].

\bibitem{Fidkowski:2003nf}
L.~Fidkowski, V.~Hubeny, M.~Kleban and S.~Shenker, \emph{{The Black hole
  singularity in AdS / CFT}},
  \href{https://doi.org/10.1088/1126-6708/2004/02/014}{\emph{JHEP} {\bfseries
  02} (2004) 014} [\href{https://arxiv.org/abs/hep-th/0306170}{{\ttfamily
  hep-th/0306170}}].

\bibitem{Festuccia:2005pi}
G.~Festuccia and H.~Liu, \emph{{Excursions beyond the horizon: Black hole
  singularities in Yang-Mills theories. I.}},
  \href{https://doi.org/10.1088/1126-6708/2006/04/044}{\emph{JHEP} {\bfseries
  04} (2006) 044} [\href{https://arxiv.org/abs/hep-th/0506202}{{\ttfamily
  hep-th/0506202}}].

\bibitem{Grinberg:2020fdj}
M.~Grinberg and J.~Maldacena, \emph{{Proper time to the black hole singularity
  from thermal one-point functions}},
  \href{https://doi.org/10.1007/JHEP03(2021)131}{\emph{JHEP} {\bfseries 03}
  (2021) 131} [\href{https://arxiv.org/abs/2011.01004}{{\ttfamily
  2011.01004}}].

\bibitem{Rodriguez-Gomez:2021pfh}
D.~Rodriguez-Gomez and J.~G. Russo, \emph{{Correlation functions in finite
  temperature CFT and black hole singularities}},
  \href{https://doi.org/10.1007/JHEP06(2021)048}{\emph{JHEP} {\bfseries 06}
  (2021) 048} [\href{https://arxiv.org/abs/2102.11891}{{\ttfamily
  2102.11891}}].

\bibitem{deBoer:2022zps}
J.~de~Boer, D.~L. Jafferis and L.~Lamprou, \emph{{On black hole interior
  reconstruction, singularities and the emergence of time}},
  \href{https://arxiv.org/abs/2211.16512}{{\ttfamily 2211.16512}}.

\bibitem{Horowitz:2023ury}
G.~T. Horowitz, H.~Leung, L.~Queimada and Y.~Zhao, \emph{{Boundary signature of
  singularity in the presence of a shock wave}},
  \href{https://doi.org/10.21468/SciPostPhys.16.2.060}{\emph{SciPost Phys.}
  {\bfseries 16} (2024) 060}
  [\href{https://arxiv.org/abs/2310.03076}{{\ttfamily 2310.03076}}].

\bibitem{Ceplak:2024bja}
N.~\v{C}eplak, H.~Liu, A.~Parnachev and S.~Valach, \emph{{Black Hole
  Singularity from OPE}},  \href{https://arxiv.org/abs/2404.17286}{{\ttfamily
  2404.17286}}.

\bibitem{Singhi:2024sdr}
K.~Singhi, \emph{{Proper time to singularity and thermal correlators}},
  \href{https://arxiv.org/abs/2406.08553}{{\ttfamily 2406.08553}}.

\bibitem{Hartman:2013qma}
T.~Hartman and J.~Maldacena, \emph{{Time Evolution of Entanglement Entropy from
  Black Hole Interiors}},
  \href{https://doi.org/10.1007/JHEP05(2013)014}{\emph{JHEP} {\bfseries 05}
  (2013) 014} [\href{https://arxiv.org/abs/1303.1080}{{\ttfamily 1303.1080}}].

\bibitem{Banados:1992wn}
M.~Banados, C.~Teitelboim and J.~Zanelli, \emph{{The Black hole in
  three-dimensional space-time}},
  \href{https://doi.org/10.1103/PhysRevLett.69.1849}{\emph{Phys. Rev. Lett.}
  {\bfseries 69} (1992) 1849}
  [\href{https://arxiv.org/abs/hep-th/9204099}{{\ttfamily hep-th/9204099}}].

\bibitem{Banados:1992gq}
M.~Banados, M.~Henneaux, C.~Teitelboim and J.~Zanelli, \emph{{Geometry of the
  (2+1) black hole}},
  \href{https://doi.org/10.1103/PhysRevD.48.1506}{\emph{Phys. Rev. D}
  {\bfseries 48} (1993) 1506}
  [\href{https://arxiv.org/abs/gr-qc/9302012}{{\ttfamily gr-qc/9302012}}].

\bibitem{Brown:1986nw}
J.~D. Brown and M.~Henneaux, \emph{{Central Charges in the Canonical
  Realization of Asymptotic Symmetries: An Example from Three-Dimensional
  Gravity}}, \href{https://doi.org/10.1007/BF01211590}{\emph{Commun. Math.
  Phys.} {\bfseries 104} (1986) 207}.

\bibitem{Horowitz:2003he}
G.~T. Horowitz and J.~M. Maldacena, \emph{{The Black hole final state}},
  \href{https://doi.org/10.1088/1126-6708/2004/02/008}{\emph{JHEP} {\bfseries
  02} (2004) 008} [\href{https://arxiv.org/abs/hep-th/0310281}{{\ttfamily
  hep-th/0310281}}].

\bibitem{Penington:2019npb}
G.~Penington, \emph{{Entanglement Wedge Reconstruction and the Information
  Paradox}}, \href{https://doi.org/10.1007/JHEP09(2020)002}{\emph{JHEP}
  {\bfseries 09} (2020) 002}
  [\href{https://arxiv.org/abs/1905.08255}{{\ttfamily 1905.08255}}].

\bibitem{Almheiri:2019psf}
A.~Almheiri, N.~Engelhardt, D.~Marolf and H.~Maxfield, \emph{{The entropy of
  bulk quantum fields and the entanglement wedge of an evaporating black
  hole}}, \href{https://doi.org/10.1007/JHEP12(2019)063}{\emph{JHEP} {\bfseries
  12} (2019) 063} [\href{https://arxiv.org/abs/1905.08762}{{\ttfamily
  1905.08762}}].

\bibitem{Almheiri:2019hni}
A.~Almheiri, R.~Mahajan, J.~Maldacena and Y.~Zhao, \emph{{The Page curve of
  Hawking radiation from semiclassical geometry}},
  \href{https://doi.org/10.1007/JHEP03(2020)149}{\emph{JHEP} {\bfseries 03}
  (2020) 149} [\href{https://arxiv.org/abs/1908.10996}{{\ttfamily
  1908.10996}}].

\end{thebibliography}\endgroup
\end{document}